\documentclass[aps,twocolumn,epsfig,showpacs]{revtex4}
\usepackage{epsfig}
\usepackage{graphicx}
\usepackage{amsmath}
\usepackage{rotating}

\begin{document} % INITIALIZE - DONT CHANGE

\title{Control of Dynamical Localization}
\author{Jiangbin Gong \cite{bylineone}, Hans Jakob W\"{o}rner,  and Paul Brumer}
\affiliation{Chemical Physics Theory Group,Department of Chemistry,
University of Toronto,
 80 St. George Street
Toronto, Canada M5S 3H6}
\date{\today}

\begin{abstract}
Control over the quantum dynamics of chaotic kicked rotor systems is
demonstrated. Specifically, control over a number of quantum coherent
phenomena is achieved by a simple modification of the kicking field. These
include the enhancement of the dynamical localization length, the
introduction of classical anomalous diffusion assisted control for systems
far from the semiclassical regime, and the observation of a variety of
strongly nonexponential lineshapes for dynamical localization. The results
provide excellent examples of  controlled quantum dynamics in a system
that is classically chaotic and offer new opportunities to explore quantum
fluctuations and correlations in quantum chaos.
\end{abstract}

\maketitle

\section{Introduction}

The quantum kicked rotor (KR) and its classical analog, the standard map,
have long served as a paradigm for quantum and classical chaos
\cite{casatibook}.  Due to its atom optics realization \cite{raizenetc},
the KR has recently attracted renewed interest. The KR is also of
considerable interest in a variety of other fields such as condensed
matter physics \cite{fishmanprl,casatiprl}, molecular physics
\cite{fishman,averbukh}, and quantum information  science
\cite{facchi,georgeot}.

One well-known quantum effect in KR is ``dynamical localization" (DL)
\cite{casatibook}. That is, although a classical kicked rotor displays
unrestricted diffusive energy increase due to classical chaos, only a
finite number of rotational states are excited in the quantum dynamics,
with the quantum excitation probability versus the rotational quantum
number typically assuming a characteristic exponential lineshape. DL is a
pure quantum coherence effect and is therefore very sensitive to
decoherence. For example, previous studies have demonstrated that noise
\cite{noiseKR, cohen,graham}, nonperiodicity in the kicking sequences
\cite{casati90,Abal}, and quantum measurements \cite{facchi} can destroy
DL.

As a coherence effect, DL is also indicative of the possibility of
controlling the KR dynamics via quantum effects
\cite{ricebook,brumerreview,brumerbook}. Indeed, we recently showed that
the quantum phases describing the initial rotor quantum state can be
manipulated  to effectively control  quantum fluctuations in quantum chaos
and thus enhance or suppress quantum chaotic diffusion
\cite{gongcontrol,gongjcp} in KR. However, manipulating quantum phases in
initial states cannot change the unitary evolution operator of the system.
Hence, neither the average dynamical localization length nor the
characteristic lineshape of dynamical localization can be altered in this
way.

Motivated by interest in controlled classically chaotic quantum dynamics
\cite{gongcontrol,gongjcp,gong02}, and to gain more insights into the
nature of DL, we consider controlling DL and the associated energy
absorption via a modified kicked rotor (MKR) system, in which the phase of
the kicking field, or the timing of the kicking sequences (hence the
evolution operator), is actively manipulated. In particular, we consider
an MKR system in which the sign of the kicking potential is periodically changed, or
alternatively time delayed, after a certain number of kicks. As shown
below, such a slight modification of KR has profound effects on the
dynamics: whereas periodically changing the sign of the kicking potential does not
destroy DL, it dramatically changes the quantum diffusion dynamics of KR
as well as the nature of DL, offering entirely new opportunities for
controlling dynamics, as well as understanding quantum fluctuations and
correlations in quantum chaos in periodically kicked systems. For example,
we demonstrate that, with the sign of the kicking potential of KR periodically
changed: (i) the dynamical localization length is significantly increased
so that the energy absorption is strongly enhanced; (ii) classical
anomalous diffusion (which can be slower than quantum anomalous diffusion
under certain conditions \cite{gongpre03}) can enhance control even when
the effective Planck constant is about an order of magnitude larger than
the relevant classical phase space structures, and (iii) DL may display
strong deviations from purely exponential lineshapes.

This paper is organized as follows: In Sec. \ref{MKR} we introduce the
modified  kicked-rotor model.  We then present results in Sec.
\ref{enhancement} on the enhancement of dynamical localization length,
with qualitative explanations based upon a known result from the Band
Random Matrix theory \cite{casatiBRM,gongBRM,izrailev}. In Sec.
\ref{anomalous} we show control of DL in a different regime, where the
dynamics can be tied to a new mechanism, i.e., the creation of new
structures in classical phase space. Section \ref{non-ex} contains the
results on coherent manipulations of the lineshapes for DL.  We conclude
the paper with a brief discussion in Sec. \ref{diss}.

\section{A Modified Kicked-rotor Model}
\label{MKR}

The KR Hamiltonian is given by
\begin{eqnarray}
H^{KR}(\hat{L},\theta, t)=\hat{L}^{2}/2I+\lambda \cos(\theta)\sum_{n}\delta(t/T-n),
\end{eqnarray}
where
$\hat{L}$ is the angular momentum operator, $\theta$ is
the conjugate angle,  $I$ is the moment of inertia, $\lambda$
is the strength of the kicking field, and $T$ is the time interval between kicks.
The basis states of their Hilbert spaces are given by $|m\rangle$,
with $\hat{L}|m\rangle=m\hbar|m\rangle$.
The quantum KR map operator
for propagating from time $(N-0^{+})T$ to time  $(N+1-0^{+})T$ is given
by
\begin{eqnarray}
\hat{F}_{KR}=\exp\left[i\frac{\tau}{2}\frac{\partial^{2}}{\partial \theta^{2}}\right]
\exp[-ik\cos(\theta)],
\label{qmapkr}
\end{eqnarray}
with dimensionless parameters
$ k=\lambda T/\hbar$ and the effective Planck constant
$\tau=\hbar T/I$.
For later use we also define the dimensionless scaled rotational energy as $\tilde{E}\equiv \langle
\hat{L}^{2}\rangle\tau^{2}/2\hbar^{2}$, where $\langle\cdot\rangle$ represents the average
over the quantum ensemble.
In the $|m\rangle$ representation, $\hat{F}_{KR}$ takes the following form \cite{izrailev}:
\begin{eqnarray}
\langle m_{1}|\hat{F}_{KR}|m_{2}\rangle=\exp\left(i\frac{\tau}{2} m_{1}^{2}\right)i^{m_{1}-m_{2}}
J_{m_{1}-m_{2}}(k),
\label{bessel}
\end{eqnarray}
where $J_{m_{1}-m_{2}}(k)$ is the Bessel function of the first kind of order $(m_{1}-m_{2})$.

The classical limit of the KR quantum map, i.e., the standard map, depends only
on one parameter $\kappa\equiv k\tau$ and is given by:
\begin{eqnarray}
\tilde{L}_{N} &= &\tilde{L}_{N-1}+\kappa \sin(\theta_{N-1}); \nonumber \\
\theta_{N}&=&\theta_{N-1}+ \tilde{L}_{N},
\label{clamap}
\end{eqnarray}
where
$\tilde{L}\equiv L\tau/\hbar$ is the scaled c-number angular momentum
and $(\tilde{L}_{N}, \theta_{N})$ represents
the phase space location of a classical
trajectory at $(N+1-0^{+})T$.
For later discussion we note that for particular values of $\kappa$,
the classical map Eq. (\ref{clamap}) can generate accelerating
trajectories whose momentum increases (or decreases) linearly with time (at least on the average).  These trajectories
are called {\it transporting}  trajectories \cite{schanz}.  To see this
consider
the initial conditions: $
(\tilde{L}=2\pi l_{1}, \theta=\pm \pi/2)$
for $\kappa=2\pi l_{2}$, where $l_{1}$ and $l_{2}$ are integers.  Clearly, these phase space points
are shifted by a constant value ($\pm 2\pi l_{2}$) in $\tilde{L}$ after each iteration,
resulting in a
quadratic
increase of rotational energy.  These transporting trajectories
are rather stable insofar as they may persist
for values of $\kappa$ close to $2\pi l_{2}$ (with their average momentum
shift after each iteration oscillating around the constant value $\pm 2\pi l_{2}$),
thus giving rise to
transporting regular islands \cite{schanz}, i.e., the accelerator modes in the KR case.
If classical
trajectories are launched from the accelerator modes,  they
simply jump to other similar islands located in adjacent phase space cells.
For trajectories initially outside the  accelerator modes,
the ``stickiness'' of the boundary between
the accelerator modes and the chaotic sea induces
anomalous diffusion over the energy space, i.e., energy increases
in a nonlinear fashion, but not quadratically.  This is intrinsically different from the
case of normal chaotic diffusion in which energy increases linearly with the number of
kicks.

We introduce here a slightly modified kicked rotor (MKR) system whose Hamiltonian is given by:
\begin{eqnarray}
H^{MKR}(\hat{L},\theta, t)=\hat{L}^{2}/2I+\lambda \cos(\theta)\sum_{n}f_{M}(n)\delta(t/T-n),
\end{eqnarray}
where $f_{M}(n)$ is real,
$|f_{M}(n)|=1$, and $f_{M}(n)$ changes sign after every $M$ kicks.
That is, the only difference between KR and MKR is that
in the MKR the sign of the  kicking potential is changed after every $M$ kicks.
The effect of changing the sign of the kicking potential can be further understood in
terms of the time-evolving wave function, which can be expanded as a superposition of
$|m\rangle$ states: $\sum_{m} C_{m} \langle\theta|m\rangle$, with
the expansion coefficients $C_{m}$.
Since $\cos(\theta+\pi)=-\cos(\theta)$,
and $\sum C_{m} \langle\theta +\pi |m\rangle$=$\sum_{m} (-1)^{m}
C_{m} \langle\theta |m\rangle$,
changing the sign of the kicking potential
is seen to be equivalent to
adding a
$\pi$ phase difference between all neighboring basis sates.
Compare this now to the effect of a time delay $t_{d}=2\pi T/\tau$ between two neighboring kicks.
Due to the free evolution of the rotor any two angular momentum eigenstates
$|m\rangle$ and $|m+1\rangle$ will
acquire in time $t_{d}$ an
additional relative quantum phase given by $\exp\{i\tau t_{d}[(m+1)^{2}-m^{2}]/(2T)\}
=\exp[i(2m+1)\pi]=\exp(i\pi)$.
As such,
the MKR
can be also realized
by introducing the time delay $t_{d}$  after every $M$ kicks.

For times $(N-0^{+})T$ to $(N+M-0^{+})T$, the MKR quantum propagator
can be written as
\begin{eqnarray}
\hat{F}_{MKR} =
\exp\left(i\pi\frac{\partial^{2}}{\partial\theta^{2}}\right)\hat{F}_{KR}^{M}
\equiv \hat{D} \hat{F}_{KR}^{M}.
\label{mkr-map}
\end{eqnarray}
where this equation defines  $\hat{D}$ as the free evolution operator over time $t_{d}$.
Here
$\hat{F}^{M}_{KR}$ denotes $M$ applications of $\hat{F}_{KR}$.
From Eq. (\ref{mkr-map}), one sees that
 the only difference in time propagation between KR and MKR
 for every $M$ kicks is the $\hat{D}$ operator, whose matrix elements
 $\langle m_{1}|\hat{D}|m_{2}\rangle$ are
 given by
 \begin{eqnarray}
 \langle m_{1}|\hat{D}|m_{2}\rangle=(-1)^{m_{1}}\delta_{m_{1}m_{2}}.
 \label{dmatrix}
 \end{eqnarray}
 Thus we have
 \begin{eqnarray}
 \langle m_{1}|\hat{F}_{MKR}|m_{2}\rangle
 =(-1)^{m_{1}}\langle m_{1}|\hat{F}_{KR}^{M}|m_{2}\rangle.
 \label{d2}
 \end{eqnarray}
 Note also that
 the classical limit of the MKR quantum map [Eq. (\ref{mkr-map})] is given  by
 \begin{eqnarray}
 \tilde{L}_{N+1}&=&\tilde{L}_{N}+\kappa f_{M}(N)\sin(\theta_{N}), \nonumber \\
  \theta_{N} &= &\theta_{N}+ \tilde{L}_{N+1}.
  \label{cla-MKR}
 \end{eqnarray}

With respect to model systems in the literature, the MKR here can be
regarded as a specific realization of the so-called generalized
kicked-rotor model,  which was first introduced in Ref. \cite{danapre}, in
the context of quantum anti-resonance. However, it is very different from
the amplitude-modulated kicked rotor systems
\cite{shepelyansky2,casati90} previously studied because (i) the kicking
field strength here remains constant, and therefore any interesting
results arise from pure phase modulations; and (ii) as shown below, the
dynamical localization is significantly altered, but not destroyed.

\section{Enhanced Dynamical Localization Length}
\label{enhancement}
As an example, consider the case of $k=4.0$, $\tau=2.0$, and $M=50$, whose
classical limit for both the KR and MKR is fully chaotic and displays
normal chaotic diffusion. We demonstrate below that the dynamical
localization (DL) and the related energy absorption associated with
$\hat{F}_{MKR}$ are dramatically enhanced over that associated with
$\hat{F}_{KR}^{M}$.

 \begin{figure}[ht]
    \begin{center}
     \epsfig{file=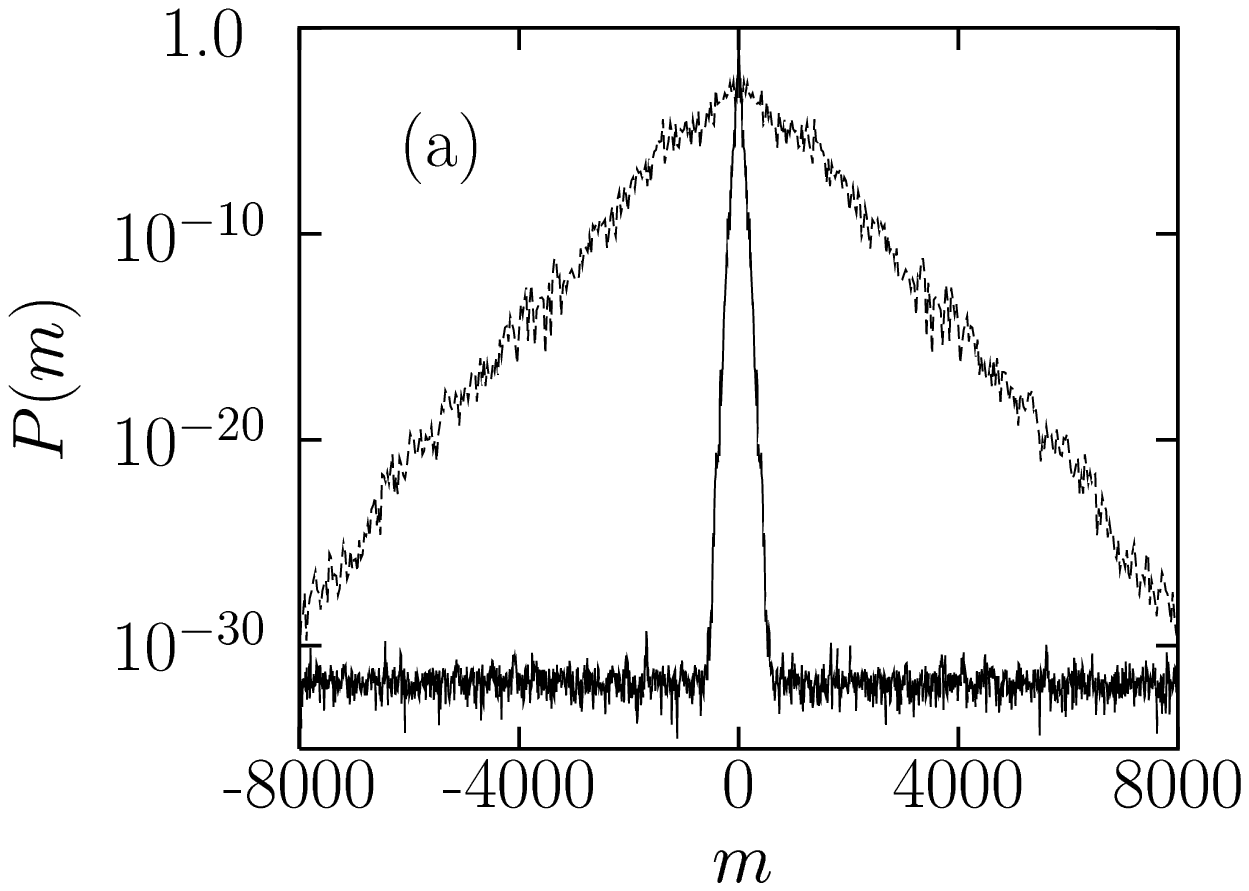,width=6cm}
     \vspace{0.5cm}

     \epsfig{file=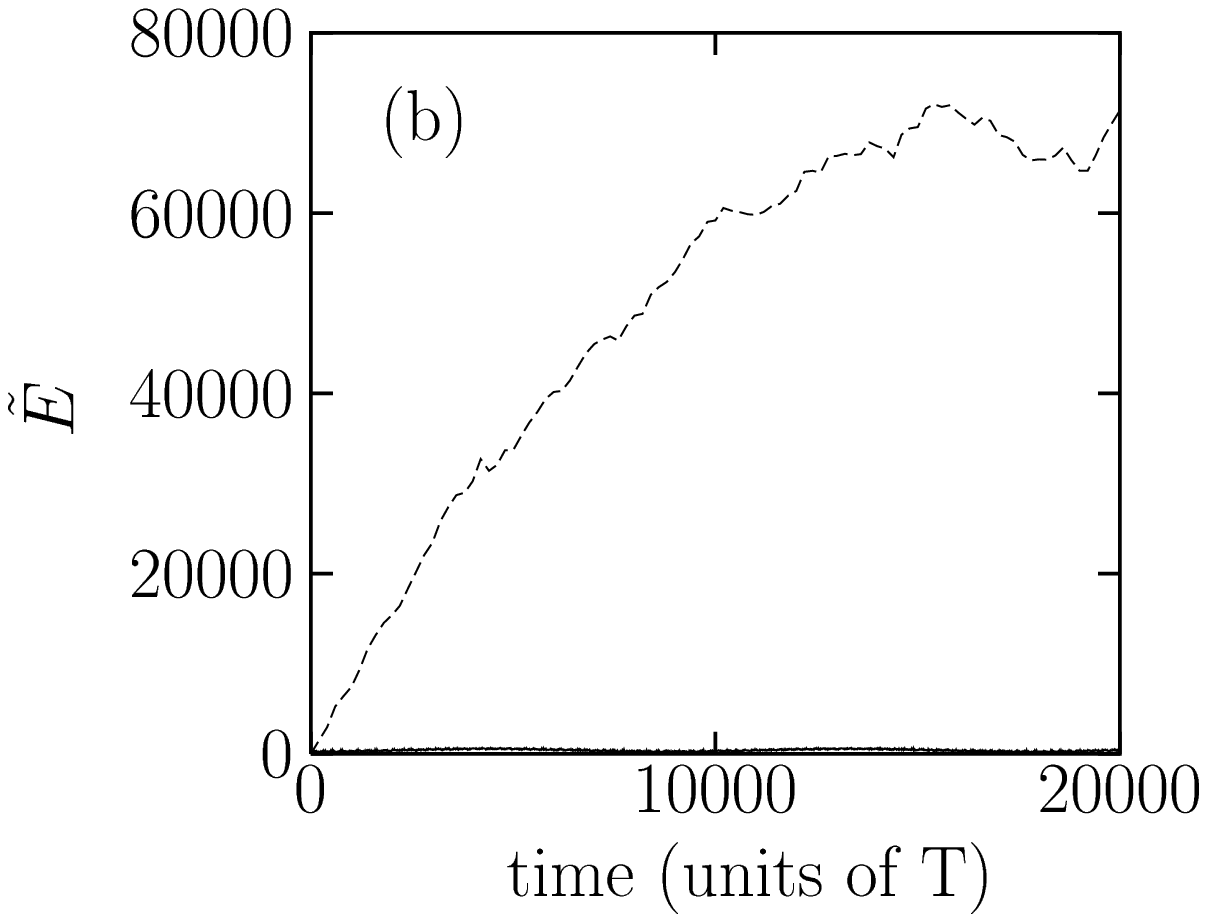,width=6cm}
      \end{center}
 \caption{Phase control of dynamical localization achieved by changing the sign of the kicking potential after every 50
kicks. $\kappa=4.0$, $\tau=2.0$, and the initial state is $|0\rangle$.
(a) A comparison between KR (the narrow lineshape, solid line) and MKR (the broad lineshape, dashed line)
in terms of the probability $P(m)$
of finding the system in the state $|m\rangle$ after  $4\times10^{5}$ kicks.
(b)
The time dependence of the dimensionless scaled rotational energy $\tilde{E}$ in each case of
KR (solid line) and MKR (dashed line).
Note that the solid line lies very closely to the $\tilde{E}=0$ axis.
}
 \label{hans-fig1}
 \end{figure}

Figure \ref{hans-fig1}a displays the angular momentum distribution $P(m)$
after $4\times 10^{5}$ kicks, starting with the initial state $|0\rangle$,
for both KR and the MKR with $M=50$. The exponential line shape of $P(m)$
shown in Fig. \ref{hans-fig1}a indicates that DL occurs in both cases. By
fitting $P(m)$ with exponentials $P(m)\sim \exp[-|m|/l_{KR}]$ and
$P(m)\sim \exp[-|m|/l_{MKR}]$ for KR and MKR, respectively, one obtains
that the dynamical localization length $l_{MKR}\sim 140.0$ is
significantly larger than $l_{KR}\sim 7.0$. This clear difference in
dynamical localization length is also reflected in the energy absorption
shown in Fig. \ref{hans-fig1}b. In particular, while the energy absorption
of KR saturates after a few kicks, the MKR system continues to absorb
energy in a more or less linear manner  for as long as $10^{4}$ kicks.
Enhancement of dynamical localization length and energy absorption is also
observed for other values of $M$,  and for a wide range of parameters $k$
and $\tau$.

As is well known, the DL of KR can be traced back to the localization
properties of the eigenstates of the quantum map operator [Eq.
(\ref{qmapkr})]. Note first that, due to the rapid decay of
$J_{m_{1}-m_{2}}(k)$ with increasing $|m_{1}-m_{2}|$ and the pseudo-random
nature of the function $\exp(i\tau m_{1}^{2}/2)$ in $m_{1}$, the quantum
map operator $\hat{F}_{KR}$ in general assumes a band structure and
behaves in a pseudo-random manner in the $|m\rangle$ representation.
Hence,  below we qualitatively  consider the MKR results in terms of a
well-known feature from Band-Random-Matrix theory
\cite{casatiBRM,gongBRM,izrailev}, namely, that the larger the band width
of the quantum map operator the larger the dynamical localization length.

Note first that the matrix $\langle m|\hat{F}_{KR}|m'\rangle$ is
pseudo-random. However, we do not expect $\langle
m|\hat{F}_{KR}^{M}|m'\rangle$ to be a pseudo-random matrix of the same
type since multiplying $\hat{F}_{KR}$ by itself $M$ times is expected to
establish correlations between the matrix elements $\langle
m_{1}|\hat{F}_{KR}^{M}|m_{2}\rangle$. Nevertheless, we do assume that the
matrix $\langle m|\hat{F}_{MKR}|m'\rangle$ is banded and pseudo-random
since the eigenstates of $\hat{F}_{MKR}$ have no simple connection  with
those of $\hat{F}_{KR}$. Consider now an arbitrary matrix element $\langle
m_{1}|\hat{F}_{KR}^{M}|m_{2}\rangle$.  Due to the quantum diffusive
dynamics within the $M$ kicks,  the $\langle
m_{1}|\hat{F}_{KR}^{M}|m_{2}\rangle$ with $|m_{1}- m_{2}|>>1$ should be
much greater than $\langle m_{1}|\hat{F}_{KR}|m_{2}\rangle$ (nevertheless,
both of them can be very small). According to  Eq. (\ref{d2}), this
implies that the matrix element $\langle m_{1}|\hat{F}_{MKR}|m_{2}\rangle$
is also much greater than $\langle m_{1}|\hat{F}_{KR}|m_{2}\rangle$.  In
this sense, we expect that the $\langle m|\hat{F}_{MKR}|m'\rangle$ matrix
should display a wider band than does $\langle m| \hat{F}_{KR}|m'\rangle$.

The band width of the matrix $\langle m|\hat{F}_{KR}|m'\rangle$ can be
defined by choosing a cut-off value for its matrix elements. One
traditional choice is $|m_{1}-m_{2}| \sim k$. In this case, $\langle
m_{1}|\hat{F}_{KR}|m_{2}\rangle$ at the boundary of the band is on the
order of $J_{k}(k)$, a number which is sufficiently small. However, this
cut-off value, if applied to the matrix $\langle
m|\hat{F}_{MKR}|m'\rangle$, would yield almost the same band width as
$\hat{F}_{KR}$.  Hence, quantitatively characterizing the band structure
of the matrix $\langle m|\hat{F}_{MKR}|m'\rangle$ is subtle, since the
very small matrix elements $\langle m_{1}|\hat{F}_{MKR}|m_{2}\rangle$ must
play an important role in enhancing the dynamical localization length of
MKR.

 \begin{figure}[ht]
 \begin{center}
\epsfig{file=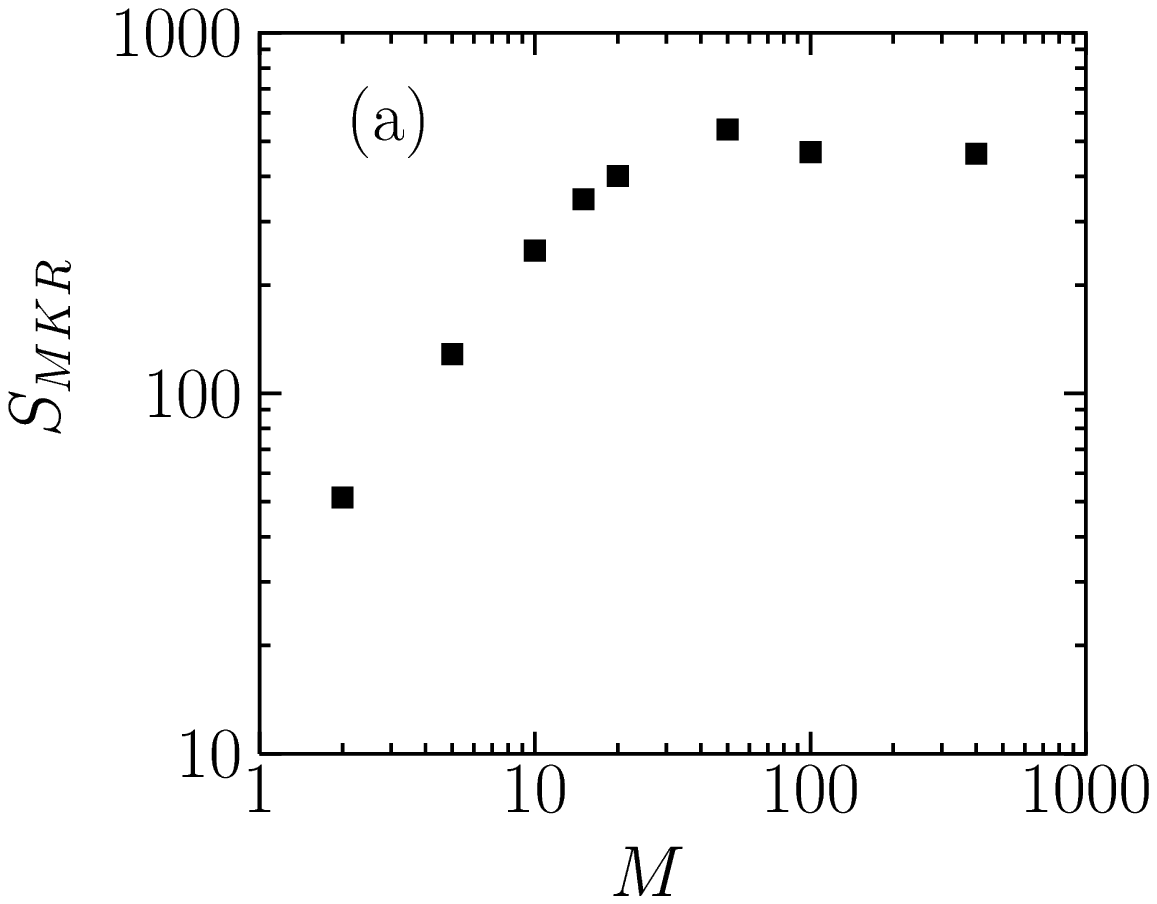,width=6cm}
\vspace{0.5cm}

 \epsfig{file=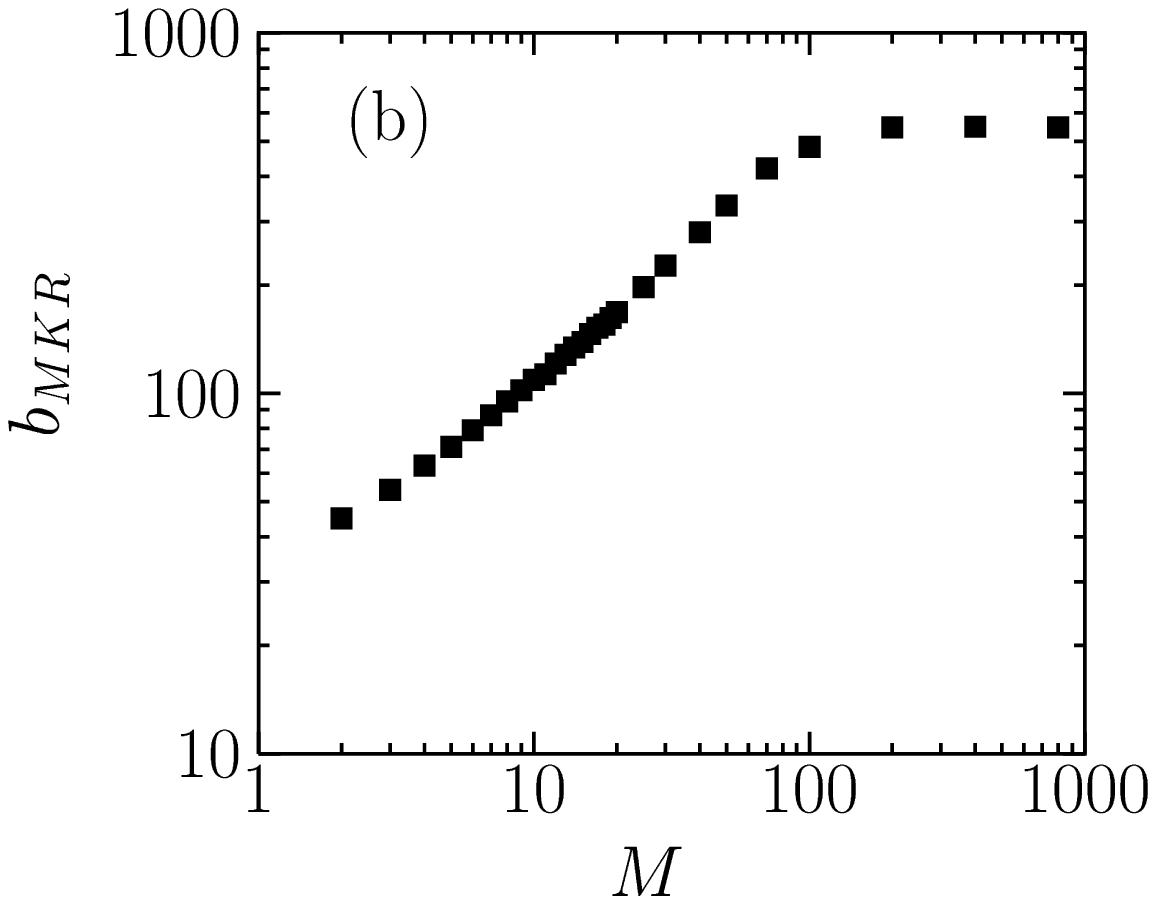,width=6cm}
 \end{center}
  \caption{(a) The $M$-dependence of the dynamical localization length of MKR,
characterized by
the Shannon entropy $S_{MKR}$ averaged over all approximate eigenfunctions of the MKR propagator $\hat{F}_{MKR}$.
(b) The $M$-dependence of the band width $b_{MKR}$ associated with the MKR propagator
$\hat{F}_{MKR}$,
defined by a cut-off value (as small as $10^{-20}$) for its matrix elements
$\langle m_{1}|\hat{F}_{MKR}|m_{2}\rangle$.
}
   \label{hans-fig2}
    \end{figure}

To gain more insight we computationally examined the $M$-dependence of the
dynamical localization length of the MKR. Specifically, we numerically
diagonalized $\hat{F}_{MKR}$, where  each matrix is generated using
$2^{14}$ basis states and is then truncated at dimension $d$ chosen below.
We characterize the average dynamical localization length by the Shannon
entropy  $S_{MKR}$ \cite{sentropy} (note that  the Shannon entropy is
simply proportional to the dynamical localization length \cite{sentropy})
averaged over all approximate eigenstates $|\phi_{j}\rangle$ of
$\hat{F}_{MKR}$, i.e.,
\begin{eqnarray}
S_{MKR}=\frac{2}{\alpha d}\sum_{j=1}^{d}\exp\left[-\sum_{m=-d/2}^{d/2}|\langle m|\phi_{j}\rangle|^{2}
\ln |\langle m|\phi_{j}\rangle|^{2}\right],
\end{eqnarray}
where the constant $\alpha$ equals 0.96, and $d$ is chosen to be 2700.
The results for the $M$-dependence of $S_{MKR}$, for
$M=2$ to $M=400$ is shown in a $\log-\log$ plot in Fig. \ref{hans-fig2}a.
$S_{MKR}$ is seen to behave initially as a smooth increasing
function of $M$, and then
to saturate at $M\sim 50$. To explain the results from the perspective of Band-Random-Matrix
theory,
we choose a cut-off value for matrix elements
$\langle m_{1}|\hat{F}_{MKR}|m_{2}\rangle$ so as to define the band width
$b_{MKR}$.
Interestingly, we find that
this cut-off value must be extremely small (roughly speaking,  at least as small as $10^{-10}$)
in order that
the $M$-dependence of $b_{MKR}$ resembles that of $S_{MKR}$. 
For example,
Fig. \ref{hans-fig2}b displays $b_{MKR}$ as a function of $M$ for a cut-off value of $10^{-20}$.
The evident similarities between Fig. \ref{hans-fig2}a and Fig. \ref{hans-fig2}b suggest
(i) that we can indeed qualitatively
explain the enhanced dynamical localization length in terms of
the Band-Random-Matrix theory, and (ii) that even extremely small quantum fluctuations in
the values of the matrix elements of the MKR quantum map operator  affect
its dynamical localization length.

In particular,
comparing Fig. \ref{hans-fig2}a with Fig. \ref{hans-fig2}b, one sees that
the saturation behavior of $S_{MKR}$ for large $M$ is consistent with the saturation behavior
of $b_{MKR}$. The latter reflects the
saturation of the matrix elements $\langle m_{1}| \hat{F}_{MKR}|m_{2}\rangle$
and therefore
the matrix elements $\langle m_{1}| \hat{F}_{KR}^{M}|m_{2}\rangle$. As such,
we infer that the saturation behavior of $S_{MKR}$
is simply a result of DL in KR.

Detailed studies on numerous other cases with varying $k$ and $\tau$ show
that the above result, i.e., $S_{MKR}$ first increases with $M$ and then
saturates, is quite general, as long as the system has a completely
chaotic classical limit and quantum correlations are insignificant. On the
other hand, if $S_{MKR}$ behaves differently, then there are two possible
origins: either the system is in the deep quantum regime or there are
non-negligible regular islands in the classical phase space. For example,
in the next section we show cases in which the energy absorption in the
MKR with $M=2$ is appreciably larger than that in the MKR with $M=3$. In
these cases one can obtain even more significant changes in DL.

\section{Classical Anomalous Diffusion Assisted Control}
\label{anomalous}
In the previous section we studied cases where both KR and MKR essentially
have a fully chaotic classical limit.  However, as shown below,  the MKR
can also display new non-chaotic classical phase space structures that are
absent in KR. In particular, regular islands with very interesting
transporting properties can be induced in MKR. In such cases one can
achieve even more dramatic alteration of the DL than that shown above.

Consider first the classical MKR map Eq. (\ref{cla-MKR}) for $M=2$.
Interestingly, in this case there exists transporting trajectories that
are different from those in KR. In particular, we have previously
observed \cite{gongpre03} that for $\kappa=(2l_{2}+1)\pi$, trajectories
emanating from $\tilde{L}=(2l_{1}+1)\pi, \theta=\pm \pi/2$ will be shifted
by a constant value [$\pm (2l_{2}+1)\pi$] in $\tilde{L}$ after each kick.
This observation suggests that new transporting regular islands that
differ from the accelerator modes in KR can be created by changing the sign of the
kicking potential after every two kicks.  This is confirmed in our
extensive numerical studies, both here and in Ref. \cite{gongpre03}.

Note that in our previous
work \cite{gongpre03}, we were most interested in the quantum-classical comparison in
anomalous diffusion and considered relatively small effective Planck constants.
In that case we found that quantum anomalous diffusion
induced by the new transporting regular islands can be much faster than
the underlying classical anomalous diffusion.
Here, to make a closer connection to atom optics experiments we consider
larger $\tau\sim 1.0$. In these cases the effective
Planck constant is about an order of magnitude
larger than the area of the phase space structures associated with classical anomalous diffusion.
Intuitively, one would anticipate that such transporting regular islands
are too small to be
relevant to the quantum dynamics.
Surprisingly, this intuition is incorrect, as shown below.

 \begin{figure}[ht]
 \begin{center}
 \epsfig{file=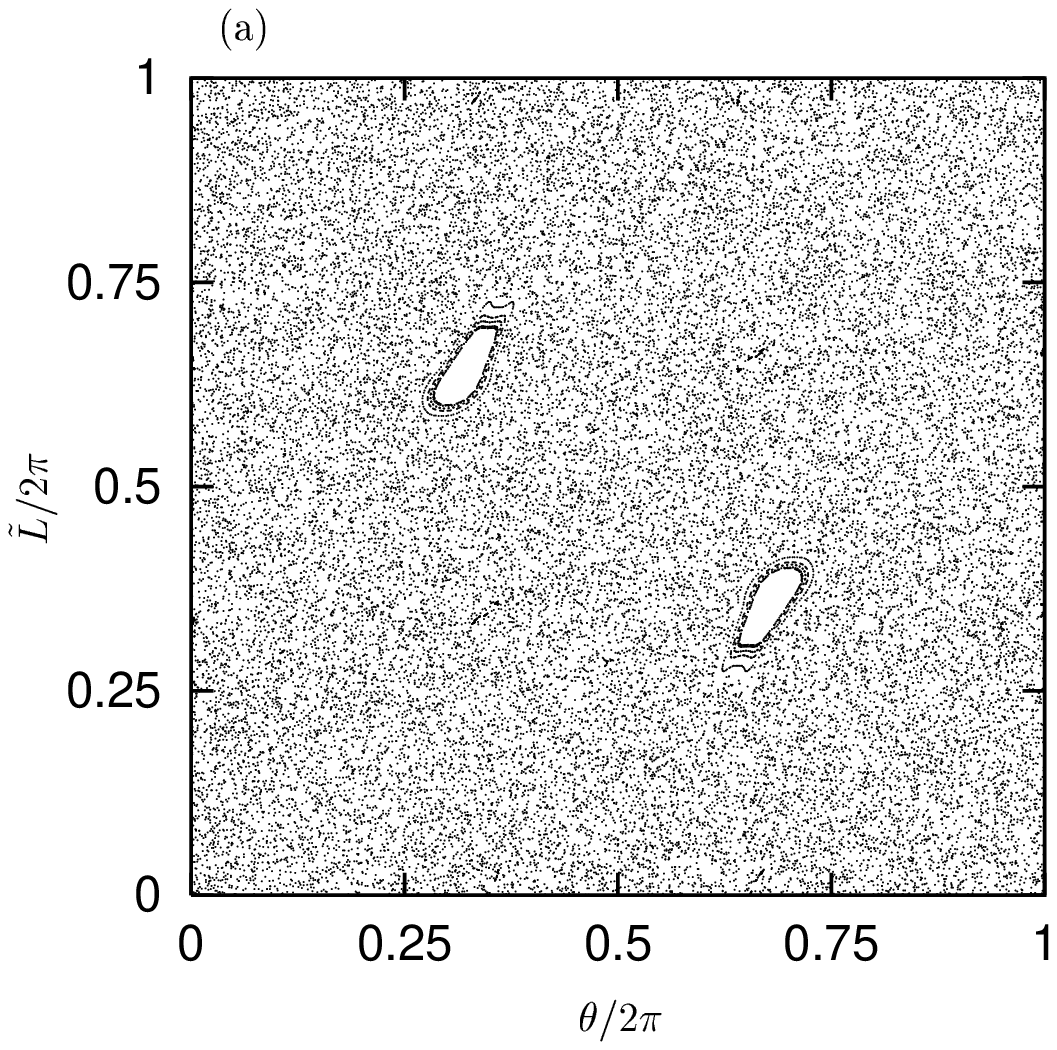,width=6cm}
    
 \vspace{0.2cm} \epsfig{file=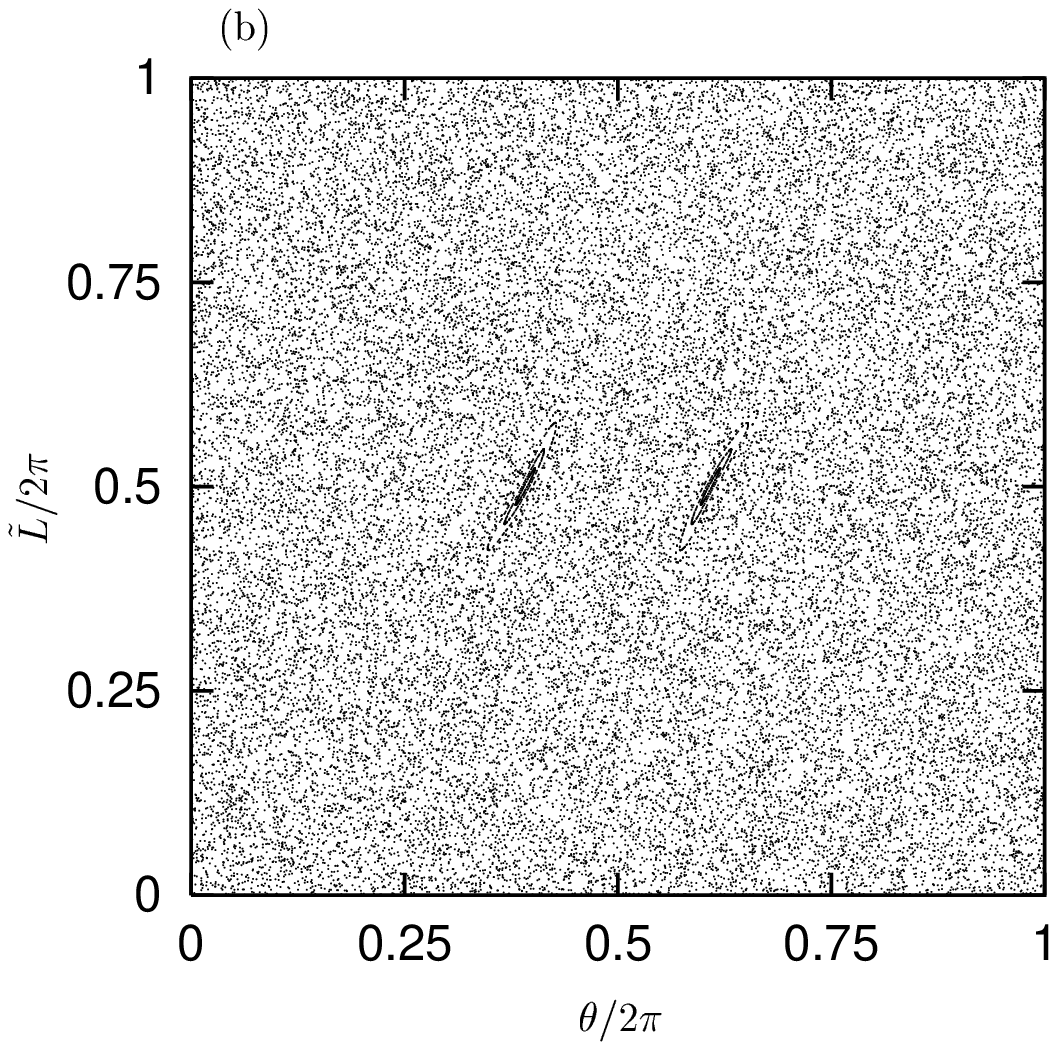,width=6cm}
 \end{center}
 \caption{Classical phase space structures of (a) the standard map and
 (b) the map of Eq. (\ref{cla-MKR}) with $M=2$, in the case of $\kappa=5.0$. All variables are
 in dimensionless units.
Note that the small regular islands seen in panel (b) are
transporting while those in panel (a) are not.}
 \label{mapkappa5}
 \end{figure}

\begin{figure}[ht]
 \begin{center}
  \epsfig{file=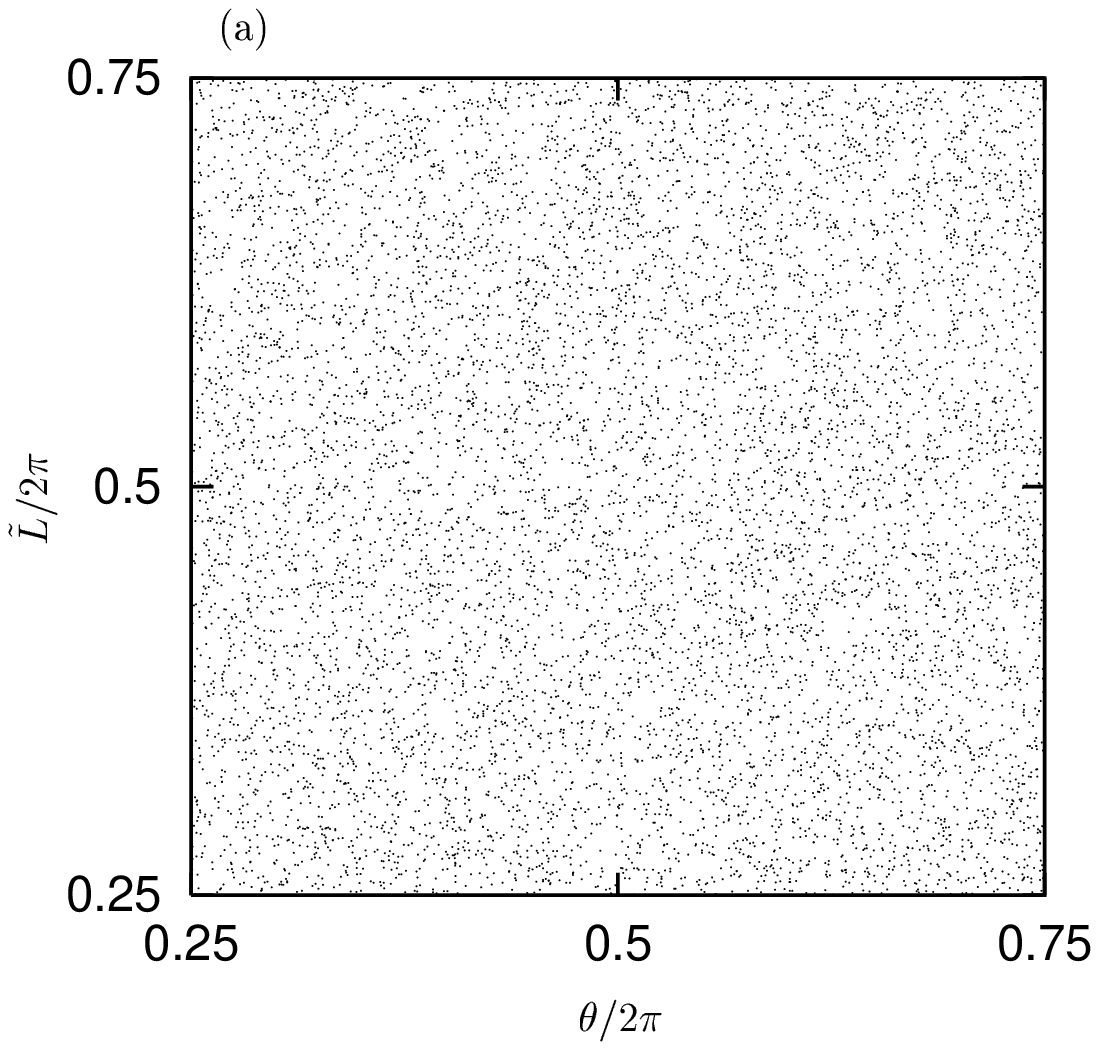,width=6cm}

  \vspace{0.2cm} \epsfig{file=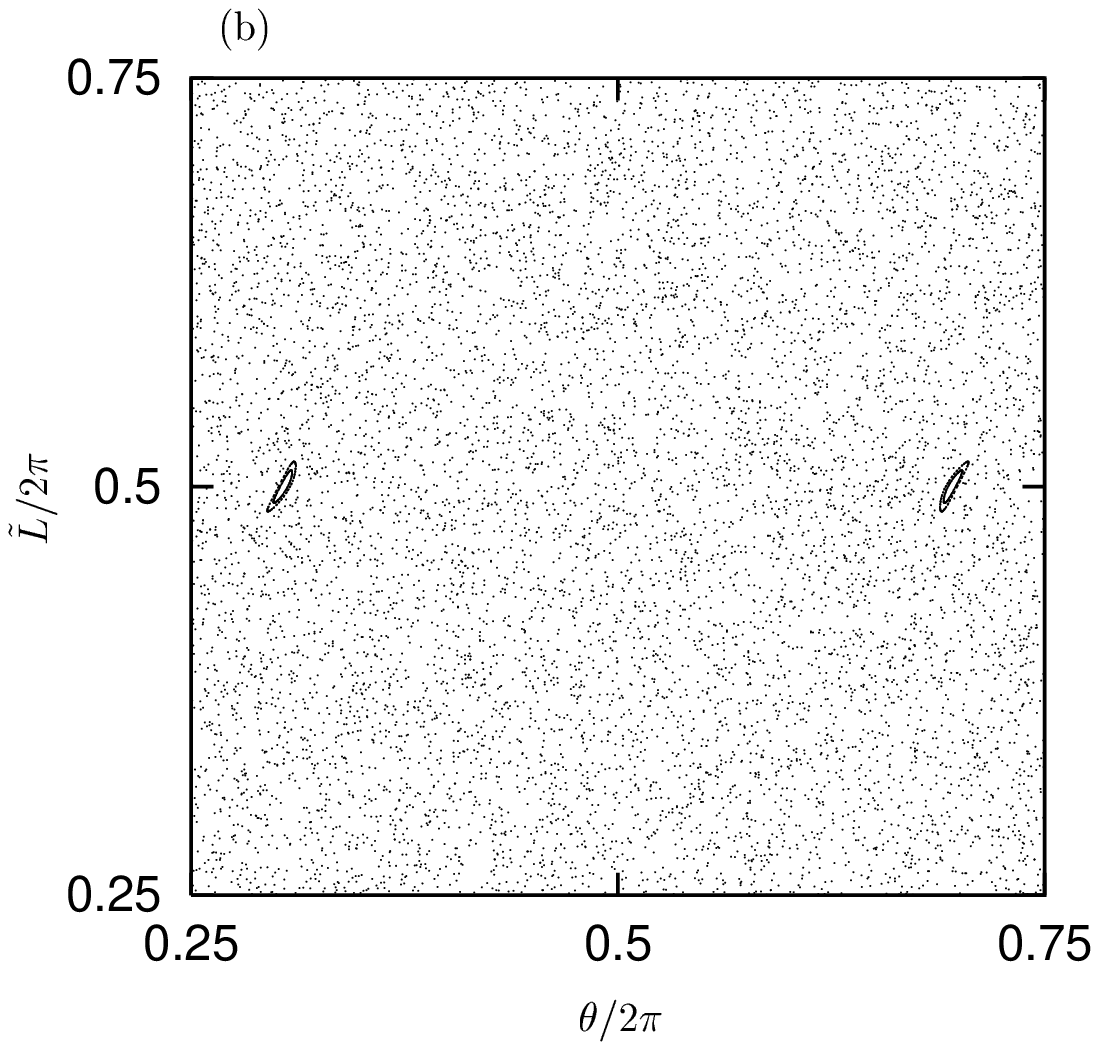,width=6cm}
  \end{center}
\caption{Same as in Fig. \ref{mapkappa5} except $\kappa=10.0$.
Note that the small regular islands seen in panel (b) are
transporting.
}
\label{mapkappa10}
 \end{figure}

To be more specific, consider first the case of  $\kappa=5.0$.  Figure \ref{mapkappa5}
displays the classical phase space structures of both KR
and the MKR with $M=2$. While the regular islands seen in Fig. \ref{mapkappa5}a (the KR case) are not
transporting (that is, the momentum of the trajectories launched from these islands
is bounded and oscillates periodically),
 a simple computation reveals that
the small islands  seen in  Fig. \ref{mapkappa5}b
in the MKR case
are transporting. That is, classical trajectories launched from the right (left) transporting regular
island
shown in Fig. \ref{mapkappa5}b have
their momentum shifted by $\pi$ ($-\pi$) on the average after each kick,
indicating that these islands originate from the
marginally stable point $\tilde{L}=(2l_{1}+1)\pi, \theta=\pm \pi/2$ with $\kappa=\pi$.
Hence, in this case phase manipulation in going from KR to the
MKR has both destroyed the nontransporting
regular islands of KR and induced new transporting regular islands.
Consider a second case with  $\kappa=10.0$. The corresponding classical
phase space structures are shown in  Fig. \ref{mapkappa10}a (KR) and Fig. \ref{mapkappa10}b
(MKR)
(Note that, to clearly display the transporting regular islands,
only a part of one phase space cell is shown here).
While there are hardly any regular islands seen in Fig. \ref{mapkappa10}a,
two small transporting regular islands are seen in Fig. \ref{mapkappa10}b.  The average momentum
shift for each kick associated with these two islands is found to be $\pm 3\pi$,  consistent with
the fact that $\kappa=10.0$ is close to $3\pi$.

Consider now the quantum dynamics of these systems. There have been only a
few studies on the quantum dynamics of delta-kicked systems where
classical chaos coexists with transporting regular islands. Of particular
relevance is the previous result that the accelerator modes of KR enhance
deviations from the normal DL behavior in KR \cite{sundaram,zaslavsky02},
even for systems far from the semiclassical limit. Since MKR displays new
transporting islands,  we therefore anticipate that DL may be strongly
affected by modifying the Hamiltonian from the KR to MKR system. This is
indeed seen below. The results are, however, counter-intuitive, since the
classical transporting regular islands created by phase manipulation of
the kicking field are found to have an area that is much smaller than the
effective Planck constant.

\begin{figure}[ht]
   \begin{center}
   \epsfig{file=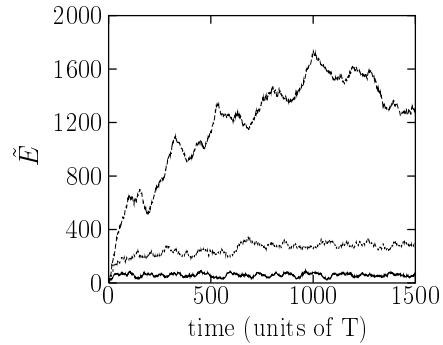, width=6cm}
     \end{center}
       \caption{The time dependence of the dimensionless scaled rotational energy $\tilde{E}$ for KR (solid
line),
for the MKR with $M=2$ (uppermost dashed line), and for the MKR with $M=3$ (middle dashed curve), with
$\tau=1.0$, $k=5.0$, and the initial state $|0\rangle$.
}
      \label{kappa5-energy}
          \end{figure}

\begin{figure}[ht]
\begin{center}
\epsfig{file=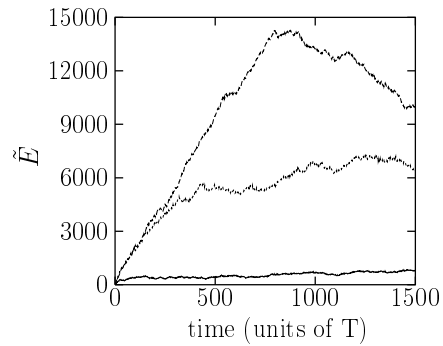, width=6cm}
 \end{center}
\caption{ Same as in Fig.  \ref{kappa5-energy} except
$\tau=1.0$, $k=10.0$.
}
 \label{kappa10-energy}
\end{figure}

For example,  for each of KR and MKR,
Figs. \ref{kappa5-energy} and \ref{kappa10-energy} display
energy absorption for the cases of
$\tau=1.0$, $k=5.0$ and $\tau=1.0$, $k=10.0$ (corresponding to $\kappa=5$ and $\kappa=10$),
respectively. Also shown is the MKR case with $M=3$, discussed below.
It is seen that the energy absorption associated with the MKR with $M=2$ (upper dashed line) is much
larger than that of KR (solid line).  Consider, for example,
$\tilde{E}$ at a specific time $t=1500T$
when the energy absorption of both KR and MKR
has clearly shown signs of saturation (e.g., the average rotational energy may decrease with time
due to statistical fluctuations).
In the first case (Fig. \ref{kappa5-energy}),
$\tilde{E}=67.7$ for KR and
$\tilde{E}=1269.7$ for MKR. In the second case (Fig. \ref{kappa10-energy}),
$\tilde{E}=798.0$ for KR
and $\tilde{E}=10049.0$ for MKR.  In both cases a control factor larger than
an order of magnitude has been achieved in going from the KR Hamiltonian to the MKR case with
$M=2$. 

This is not the case for $M=3$, $\tau=1.0$, $k=5.0$ shown in Fig.
\ref{kappa5-energy}. Here the energy absorption in the MKR with $M=3$ is
only slightly larger than in the KR and far less than in the MKR with
$M=2$. Similarly, for the case of $\tau=1.0$, $k=10.0$ shown in Fig.
\ref{kappa10-energy}, although energy absorption in the MKR with $M=3$ is
much enhanced (compared with KR), it is still not as significant as in the
MKR with $M=2$. This indicates, as confirmed by directly examining $P(m)$
(not shown) after saturation, that for both cases the dynamical
localization length of the MKR with $M=3$ is no larger than that of the
MKR with $M=2$, contrary to what is observed in the previous section. This
is because the underlying classical dynamics of MKR here is not completely
chaotic, i.e., the classical dynamics displays characteristics of
anomalous diffusion due to transporting regular islands in phase space.
Thus one can expect statistical deviations from the Band Random Matrix
theory, used previously to relate $M$ to the extent of control.

These results emphasize that the control mechanism here is uniquely based
upon the transporting regular islands created by our control scenario.
This is further supported by the lineshape for DL, which can be strongly
nonexponential, as discussed below.

\section{Nonexponential Lineshapes for Dynamical Localization}
\label{non-ex}

It was pointed out more than two decades ago \cite{fishman} that
the DL of KR can be mapped onto the problem of Anderson localization in
disordered systems.  In particular,
an exactly soluble case of disorder in tight-binding models, i.e., the Lloyd model
\cite{Lloyd}, suggests that dynamical localization  should assume an
exponential line shape, at least on the average.
This has been confirmed by numerous computational studies on KR.
For example,  Fig. \ref{hans-fig1}a clearly demonstrates, for both KR and MKR,
that the distribution function $P(m)$ can be fit beautifully with an exponential
function with a
characteristic localization length.

However, the Hamiltonian nature of the KR and MKR implies that there always exist
some subtle quantum phase correlations in the quantum dynamics.  Hence,
in addition to some universal properties of DL,
the DL lineshape can display
rich non-universal
properties, e.g., the system may display nonexponential dynamical localization.
Nonexponential dynamical localization has been previously observed in KR
but its origins are still poorly understood
\cite{satija,sundaram}.

Here, we demonstrate that the MKR with $M=2$ can display strongly
nonexponential line shapes for DL, rarely seen in KR.  We focus on the MKR
with $M=2$ since the classical MKR with $M=2$ has transporting regular
islands that are absent in KR. Second, transporting regular islands may
induce large fluctuations in DL \cite{sundaram}. However, we examine below
cases with connections to anomalous diffusion as well as those without
clear connections to anomalous diffusion. We have also studied other
versions of MKR with $M\ne 2$, and have found that nonexponential
lineshapes for DL in the latter case are much less common than in the
$M=2$ case.

  \begin{figure}[ht]
   \begin{center}
     \epsfig{file=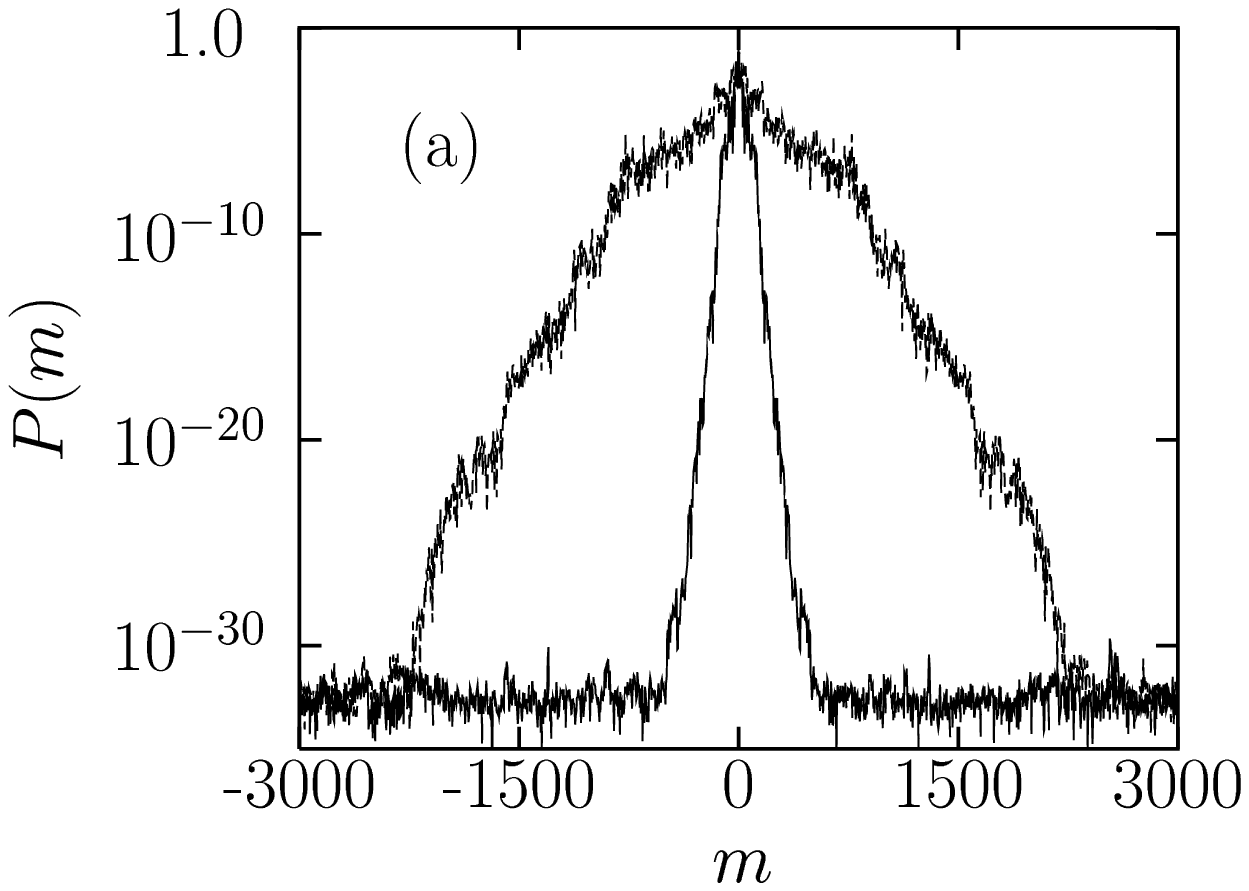,width=4cm}
       \epsfig{file=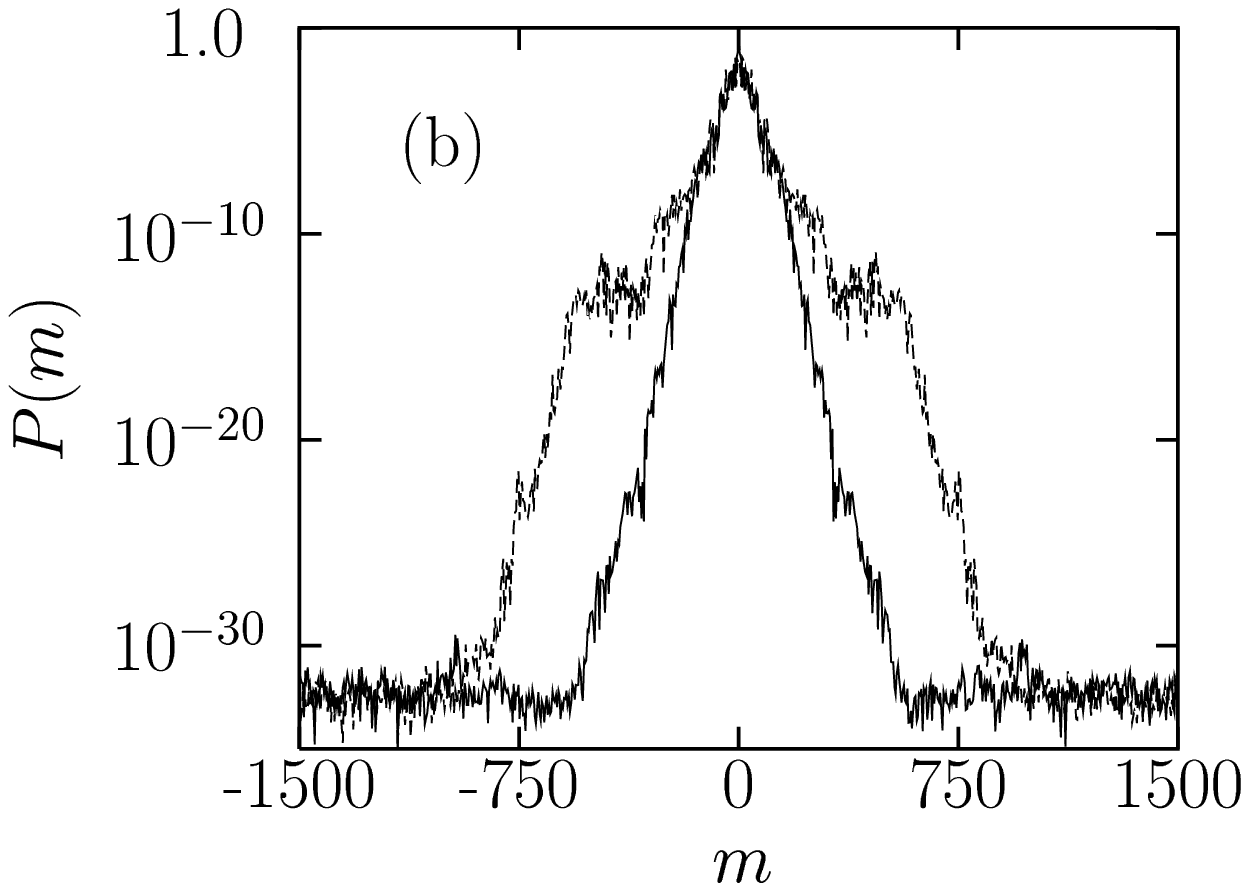,width=4cm}

       \epsfig{file=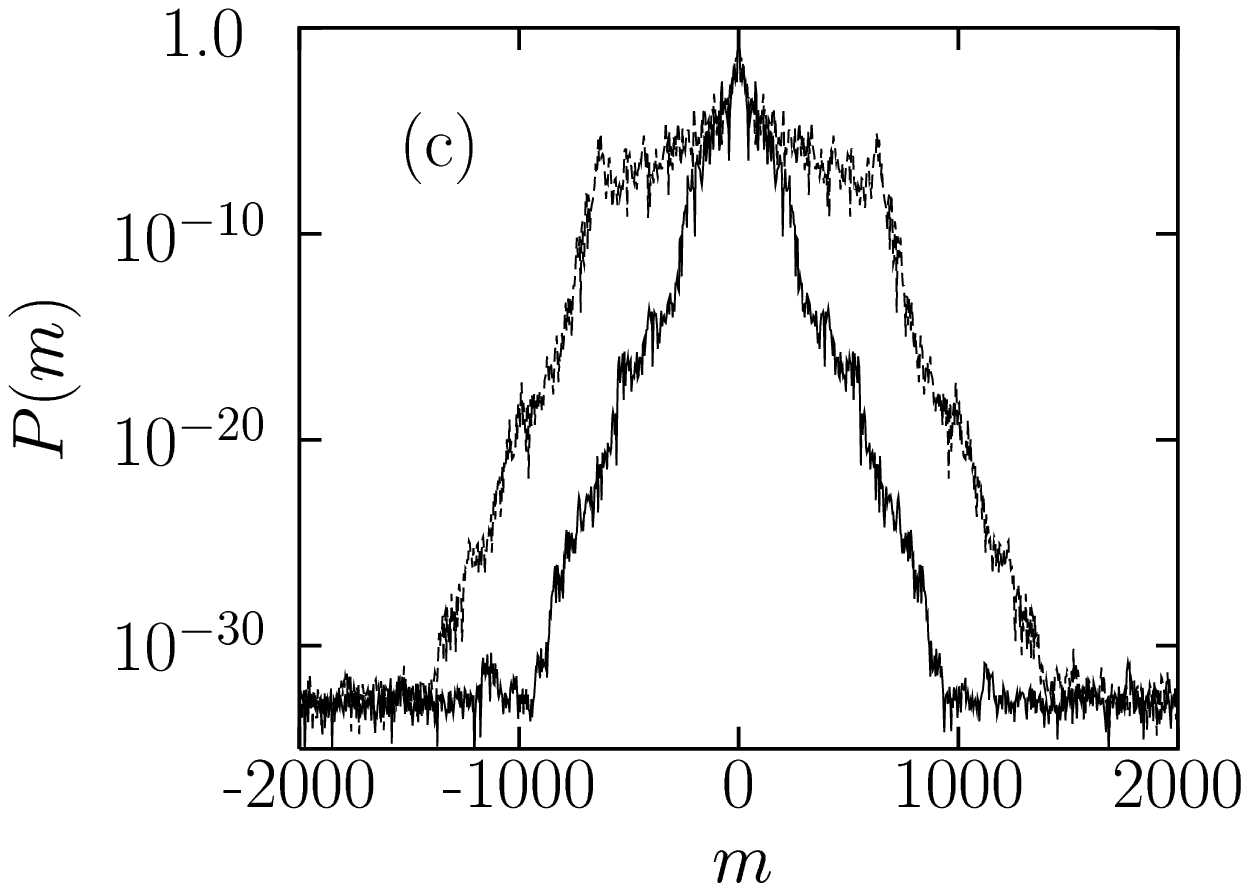,width=4cm}
          \hspace{0.2cm} \epsfig{file=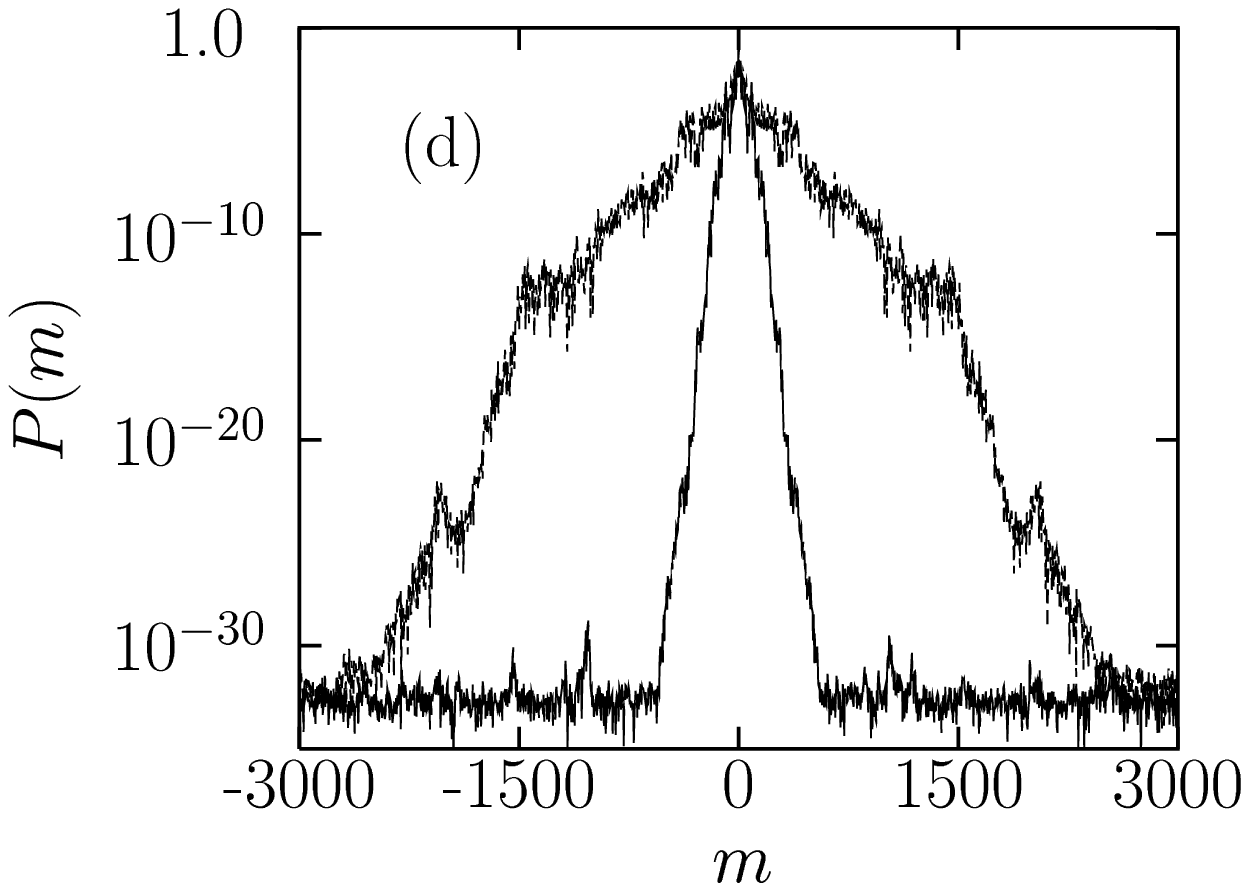,width=4cm}
     \end{center}
     \caption{Four examples of nonexponential lineshapes for dynamical localization
in the MKR with $M=2$, shown in terms of  the probability $P(m)$
of finding the system in the state $|m\rangle$ after $8000$ kicks, with the initial state
$|0\rangle$. In each case the broad lineshape is associated with the MKR,
and for the purpose of comparison,  the narrow lineshape of the analogous KR
is also shown.
(a) $\tau=1.0$, $k=5.0$, (b)  $\tau=2.0$, $k=5.0$, (c)  $\tau=1.0$, $k=5.7$,
and (d) $\tau=2.0$, $k=6.0$.
}
     \label{pdis}
      \end{figure}

Figure \ref{pdis} compares nonexponential lineshapes for DL in the MKR
(upper dashed lines) to the analogous exponential lineshapes for DL in  KR
(solid lines), for four different values of $k$ and $\tau$.  The
lineshapes are obtained by propagating the quantum dynamics for 8000 kicks
from the initial state $|0\rangle$, and will remain essentially the same
for longer propagation times \cite{note}. The huge difference between the
KR and MKR lineshapes is striking. As shown in Fig. \ref{pdis}, in the MKR
case, $P(m)$ plotted in the $\log$ scale displays structures that are far
from a purely exponential lineshape.  For example, one sees that the
initial exponential decay rate of  $P(m)$ with $|m|$ is considerably
smaller than its large-$|m|$ exponential decay rate, suggesting that
multiple characteristic lengths are needed to describe the DL of the MKR.
Also seen is that the difference in $P(m)$ between KR and MKR can be as
large as  ten orders of magnitude or more. Further,  Figs. \ref{pdis}b and
\ref{pdis}c indicate that the decay rate of $P(m)$ for large $|m|$ in the
MKR case is very similar to the KR case, although this is not the case in
Figs. \ref{pdis}a and \ref{pdis}d.

To the best of our knowledge, this is the first demonstration that intriguing
differences in the lineshapes for DL can be created by simply changing the sign of
the kicking potential periodically.

The four cases of nonexponential lineshapes shown in Fig. \ref{pdis} can
be divided into two categories, based upon the properties of their
underlying classical dynamics. The classical dynamics associated with the
cases in  Figs. \ref{pdis}a and \ref{pdis}b, shown in Figs.
\ref{mapkappa5}b and \ref{mapkappa10}b, displays transporting regular
islands.  The presence of these classical structures implies an
inhomogeneous classical phase space, and, as demonstrated in the previous
section, may have a significant impact on the quantum dynamics even when
their size is much smaller than the effective Planck constant. In this
regard, the nonexponential lineshapes shown in Figs. \ref{pdis}a and
\ref{pdis}b may not be totally surprising. However, for the other two MKR
cases shown in Figs. \ref{pdis}c and  \ref{pdis}d, we did not find any
regular islands in their classical phase space even when examined on a
very fine scale, suggesting that their classical dynamics is essentially
fully chaotic.   Thus, understanding the nonexponential lineshapes shown
in Fig. \ref{pdis}c and  Fig. \ref{pdis}d will be even more challenging.

The nonexponential lineshapes for DL arise from extremely small quantum
fluctuations and residual quantum correlations in quantum chaos. To be
able to resolve the nonexponential lineshape for DL, $P(m)$ has to be
known with high precision.  For example,  the two shoulders shown in Fig.
\ref{pdis}b involve occupation probabilities $P(m)$ as small as $\sim
10^{-14}$. It is therefore not surprising that,  while it is common to
have strongly nonexponential lineshapes for DL in the MKR case, each
individual lineshape is highly sensitive to the exact value of the
effective Planck constant. For example,  for the case shown in  Fig.
\ref{pdis}b, increasing the value of $\tau$ from $2.0$ to $2.0+10^{-5}$
can completely destroy the nonexponential lineshape!  This drastic change
in the lineshape for DL even occurs without causing an obvious difference
in energy absorption behavior. Evidently then, both experimental
observations and theoretical predictions of nonexponential dynamical
localization are far from trivial and are in need of further study.

\section{Discussion and Conclusions}
\label{diss}

This paper has dealt with control of dynamical localization in kicked
rotor systems. In all cases we manipulate the external kicking field to
alter the properties of the rotor system, i.e. the distribution of
population amongst rotor energy levels after saturation as well as the
energy absorption. In particular, we have examined the effect of
introducing a reversal of the kicking field after $M$ kicks which, within
the framework of quantum mechanics, corresponds to introducing a phase
shift amongst rotor energy levels.

Two parameter regimes have been examined, one which shows enhanced DL
lengths with increasing $M$, and the other which need not. This behavioral
difference can be understood in terms of the character of the underlying
classical phase space: the former systems are completely chaotic whereas
the latter show a mixed phase space that includes transporting regular
islands. Indeed, we have found that even if the transporting islands are
tiny compared to the effective Planck constant, they still have a profound
effect on the control of the DL. Further, a comparison of the traditional
kicked rotor system to the modified kicked rotor systems shows that the
latter is much more capable of displaying nonexponential dynamical
localization. Thus, by modifying the kicking potential are able to control
the dynamical localization in the kicked rotor.

The results of this paper are relevant to two fields of study: quantum
control and kicked rotor dynamics. From the control perspective, modifying
the kicking field changes the dynamics. However, this system does not
obviously permit a picture in terms of interfering quantum pathways (the
standard view of weak field coherent control
\cite{brumerreview,brumerbook}) since (a) the kicking field is always on,
and (b) there are a multitude of interfering transitions responsible for
the observed behavior. Indeed, it is even difficult to isolate the
interfering pathways that are responsible for dynamical localization in
the simple kicked rotor, whose dynamics is easier than that of the MKR.

From the viewpoint of  kicked rotor studies in the field of quantum chaos,
this paper provides insights into the quantum dynamics in the case
displaying classical anomalous diffusion. As one of the results of this
study, we find that classical transporting regular islands can
dramatically affect the quantum dynamics even when their size is much
smaller than the effective Planck constant. Further, the MKR system
proposed in this paper provides a new model for the study of quantum
dynamics where the underlying classical chaos coexists with transporting
regular islands. By choosing proper system parameters,  we can create
transporting regular islands whose size varies from being much larger to
being much smaller than that of the accelerator modes of KR.  Encouraged
by this, we plan in the near future to further use the MKR to study
quantum tunneling between the transporting regular island and the chaotic
sea \cite{gongpre03,zaslavsky02} and between transporting regular islands,
and to study the phase space structure of quantum eigenstates
\cite{schanz}.

In the fully chaotic case, our approach suggests the need for additional
studies of dynamical localization from the 
Band-Random-Matrix theory perspective.  For example, we have
qualitatively explained the results in Sec. \ref{enhancement} in terms of
well known features of Band-Random-Matrix theory.  However, we found that
the band width of the quantum map operator had to be defined using an 
extremely small cut-off value for the matrix elements, suggesting that a
quantitative understanding of the MKR results such as  Fig.
\ref{hans-fig2} may require new models of Band-Random-Matrix ensembles.

The strongly nonexponential lineshapes for DL found in the MKR with
$M=2$ further demonstrate the need for more theoretical work on properties
of DL. In particular, our results should motivate greater interest in
characterizing and understanding nonexponential dynamical localization,
with efforts directed towards explaining why nonexponential dynamical
localization occurs for some system parameters and not for others. This is
of importance in understanding the high sensitivity of nonexponential
lineshapes for DL to the exact value of the effective Planck constant.

We have chosen the system parameters to be within the reach of current
atom optics experiments on KR \cite{raizenetc}.  Although experimental
studies of nonexponential lineshapes for DL are difficult, we believe that
it is straightforward to experimentally observe the results of Sec.
\ref{enhancement} and Sec. \ref{anomalous}. Apart from the atom optics
realization of KR and MKR, it is also interesting to consider a molecular
version of KR and MKR, i.e., diatomics periodically kicked in strong
microwave fields \cite{fishman,gongjcp}. Preliminary computational studies
\cite{Hans} confirm that directly observing quantum control of dynamical
localization in molecular rotational motion is possible, e.g., in the case
of quantum anomalous diffusion. Another promising experimental realization
of KR and MKR requires kicked particles in a square-well potential
\cite{laksh01}. Along this direction an  interesting model (which is very
different from ours) has recently been proposed for the study of classical
and quantum anomalous diffusion \cite{laksh02}.

In summary, consideration of control in classically chaotic quantum
systems is of general interest and importance to both fields of quantum
chaos and quantum control. In this paper we have demonstrated, via a
modified kicked rotor model, that dynamical localization, perhaps the best
known phenomenon in quantum chaos, can be modified over a wide range. 
The results are of both experimental and theoretical interest.

This work was supported by the U.S. Office of Naval Research and the
Natural Sciences and Engineering Research Council of Canada.
HJW was partially
supported by the Studien-stiftung des Deutschen Volkes and the Barth Fonds of ETH
Z\"{u}rich.

%% square well reference


\begin{thebibliography}{100}
\bibitem{note}We have also examined the eigenstates of the quantum MKR propagator
  and found that a significant number of eigenstates display nonexponential lineshapes
  that are similar to what is seen from propagating the initial state $|0\rangle $.
\end{thebibliography}
\end{document}